\documentclass{article}
\usepackage{spconf,amsmath,graphicx, amsfonts,url}

\def\x{{\mathbf x}}
\def\h{{\mathbf h}}
\def\y{{\mathbf y}}
\def\Q{{\mathbf Q}}
\def\K{{\mathbf K}}
\def\V{{\mathbf V}}
\def\W{{\mathbf W}}
\def\R{{\mathbb R}}
\def\Z{{\mathbf Z}}
\def\u{{\mathbf u}}
\def\head{\mathrm{head}}
\def\model{\mathrm{model}}
\def\Lblock{L_{\mathrm{block}}}
\def\Lhop{L_{\mathrm{hop}}}

\title{Transformer ASR with Contextual Block Processing}
\name{Emiru Tsunoo$^{1}$, Yosuke Kashiwagi$^{1}$, Toshiyuki Kumakura$^{1}$, Shinji Watanabe$^{2}$}
\address{
 $^1$Sony Corporation, Japan \\
 $^2$Johns Hopkins University, USA
}
\begin{document}
\ninept
\maketitle
\begin{abstract}
The Transformer self-attention network has recently shown promising performance as an alternative to recurrent neural networks (RNNs) in end-to-end (E2E) automatic speech recognition (ASR) systems.
However, the Transformer has a drawback in that the entire input sequence is required to compute self-attention.
In this paper, we propose a new block processing method for the Transformer encoder by introducing a context-aware inheritance mechanism.
An additional context embedding vector handed over from the previously processed block helps to encode not only local acoustic information but also global linguistic, channel, and speaker attributes.
We introduce a novel mask technique to implement the context inheritance to train the model efficiently.
Evaluations of the Wall Street Journal (WSJ), Librispeech, VoxForge Italian, and AISHELL-1 Mandarin speech recognition datasets show that our proposed contextual block processing method outperforms naive block processing consistently.
Furthermore, the attention weight tendency of each layer is analyzed to clarify how the added contextual inheritance mechanism models the global information.
\end{abstract}
\begin{keywords}
Speech Recognition, End-to-end, Transformer, Self-attention Network, Block Processing
\end{keywords}
\section{Introduction}
\label{sec:intro}
In speech, the phonetic events occur at a temporally local level, whereas the speaker, channel, and long-range linguistic context exist globally.
Conventional automatic speech recognition (ASR) systems, such as fully connected neural networks estimate hidden Markov model (HMM) emission probabilities from only a locally selected audio chunk \cite{bourlard94}.
The recent success of recurrent neural networks (RNNs) in acoustic modeling \cite{graves2013hybrid,weng2014recurrent,sak15} can be attributed to exploiting the global context simultaneously with local phonetic information.

End-to-end (E2E) models are currently attracting attention as methods of directly integrating acoustic models (AMs) and language models (LMs) because of their simple training procedure and decoding efficiency.  
In recent years, various models have been studied, including connectionist temporal classification (CTC) \cite{graves06, graves14, miao15}, attention-based encoder--decoder models \cite{chorowski15, chan16, lu16, zeyer2018improved}, their hybrid models \cite{kim17, watanabe17}, and RNN transducer \cite{graves12,graves13rnnt,rao17}.
With thousands of hours of speech-transcription parallel data, E2E ASR systems have become comparable to conventional HMM-based ASR systems \cite{amodei16,prabhavalkar17analysis,li17,chiu18}.

Recently, Vaswani {\it et al.} \cite{vaswani17} have proposed a new sequence model without any recurrences, called the Transformer, which showed state-of-the-art performance in a machine translation task.
It has multihead self-attention layers, which can leverage a combination of information from completely different positions of the input.
This mechanism is suitable for exploiting both local and global information.
It also has the advantage of only requiring a single batch calculation rather than the sequential computation of RNNs.
Therefore, it can be trained faster with parallel computing.
The Transformer has been successfully introduced into E2E ASR with and without CTC by replacing RNNs \cite{dong18, sperber18, salazar19, dong19, zhao19}.

However, similarly to bidirectional RNN models \cite{schuster97}, the Transformer has a drawback in that the entire utterance is required to compute self-attention, making it difficult to utilize in online recognition systems.
Also, the memory and computational requirements of the Transformer grow quadratically with the input sequence length, which makes it difficult to apply to longer speech utterances.
A simple solution to these problems is block processing as in \cite{sperber18, dong19, jaitly2015neural}.
However, it loses the global context information and degrades performance in general.

In this paper, we propose a new block processing of the Transformer encoder by introducing a context-aware inheritance mechanism.
To utilize not only local acoustic information but also global linguistic/channel/speaker attributes, we introduce an additional context embedding vector, which is handed over from the previously processed block to the following block.
We also extend a mask technique to cope with the context inheritance to realize efficient training.
Evaluations of the Wall Street Journal (WSJ), LibriSpeech, VoxForge Italian, and AISHELL-1 Mandarin speech recognition datasets show that our proposed contextual block processing method outperforms naive block processing consistently.
The attention weight tendency of each layer is also analyzed to clarify how the added contextual inheritance mechanism works.

\section{Relation with prior work}
\label{sec:prior}
The online processing of attention-based E2E ASR by window shifting using the median \cite{chorowski15, chan16online} or the monotonic energy function as in MoChA \cite{chiu2017monotonic}, parametric Gaussian attention \cite{hou17}, and a trigger mechanism for notifying the timing of the computation \cite{moritz19} has frequently been investigated recently.
Using the Transformer ASRs, several studies have investigated online processing.
Sperber {\it et al.} \cite{sperber18} pointed out that the local context plays a crucial role in acoustic modeling.
Therefore, they tested attention biasing using local block masking and Gaussian masking, where Gaussian masking was superior because it focused on varying the granularity according to the layer.
They found that not only the local context but also the long-range channel and speaker properties were also useful for acoustic modeling.
However, Gaussian masking still requires the entire input sequence.
Dong {\it et al.} \cite{dong19} introduced a chunk hopping mechanism to the CTC-Transformer model to support online recognition, which degraded the standard Transformer since it ignored the global context.

For LMs, Transformer-XL was proposed to cope with longer-sequence modeling \cite{dai19}.
The input sequence is split into fixed-length segments, and by caching the previous segments, the context is extended recurrently.
Child {\it et al.} \cite{child19} applied the Transformer to image/text/music generation with sparse attention masking, which was similar to block processing, to generate long sequences.

In this paper, we explore the Transformer architecture for application to online speech recognition.
To prevent performance degradation due to block processing, the contextual information is inherited in a simple manner.
We utilize an additional context embedding vector that is handed over from the preceding block to the following one.
The context inheritance is performed via batch process using a mask function, rather than recursively caching the whole state of the previous blocks as in \cite{dai19}.
Thus, our model can be trained efficiently with parallel computing.

\section{Transformer ASR}
\label{sec:transformer}
\begin{figure}[t]
  \centering
  \includegraphics[width=\columnwidth]{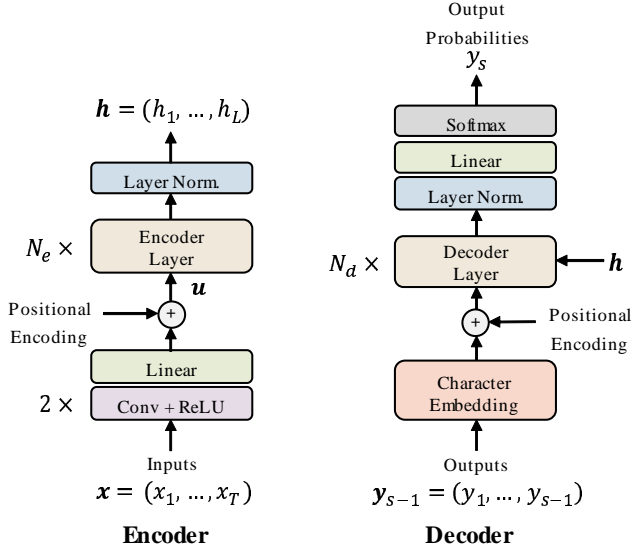}
  \vspace{-0.5cm}
  \caption{The model architecture of the Transformer ASR.}
  \label{fig:transformer}
\end{figure}

The baseline Transformer ASR follows that in \cite{dong18}, which is based on the encoder--decoder architecture in Fig.~\ref{fig:transformer}.
An encoder transforms a $T$-length speech feature sequence $\x = (x_{1},\dots,x_{T})$ to an $L$-length intermediate representation $\h = (h_{1},\dots,h_{L})$, where $L \leq T$ due to downsampling.
Given $\h$ and previously emitted character outputs $\y_{s-1} = (y_{1},\dots,y_{s-1})$, a decoder estimates the next character $y_{s}$.

The encoder consists of two convolutional layers with stride $2$ for downsampling, a linear projection layer, positional encoding, followed by $N_{e}$ encoder layers and layer normalization.
The positional encoding is a $d_{\model}$-dimensional vector defined as 
\begin{align}
    \mathrm{PE}_{(pos,2i)} = 
         \sin{(\frac{pos}{10000^{2i/d_{\model}}})} \nonumber \\
    \mathrm{PE}_{(pos,2i+1)} = 
         \cos{(\frac{pos}{10000^{2i/d_{\model}}})},
 \label{eq:position}
\end{align}
which is added to the output of the linear projection of two-layer convolutions.
Each encoder layer has a multihead self-attention network (SAN) followed by a position-wise feed-forward network, both of which have residual connections.
Layer normalization is also applied before each module.
In the SAN, the attention weights are formed from queries ($\mathbf{Q} \in \R^{t_q\times d}$) and keys ($\mathbf{K} \in \R^{t_k\times d}$), and applied to values ($\mathbf{V} \in \R^{t_v\times d}$) as
\begin{align}
    \mathrm{Attention}(\Q,\K,\V) = \mathrm{softmax}\left(\frac{\Q\K^T}{\sqrt{d}}\right)\V, \label{eq:attention}
\end{align}
where typically $d = d_{\model}/m$.
We utilized multihead attention denoted as the $\mathrm{MHD}(\cdot)$ function as follows:
\begin{align}
    & \mathrm{MHD}(\Q,\K,\V) = \mathrm{Concat}(\head_{1},\dots,\head_{m})\W_O^n,
    \label{eq:mhead}
\end{align}
where $\head_{i}$ is concatenated ($\mathrm{Concat}(\cdot)$) and linearly transformed with the projection matrix $\W_O^n$.
$m$ is the number of heads.
$\head_{i}$ is calculated with the $\mathrm{Attention}(\cdot)$ function introduced in \eqref{eq:attention} as follows.
\begin{align}
    & \head_{i} = \mathrm{Attention}(\Q\W_{Q,i}^n,\K\W_{K,i}^n,\V\W_{V,i}^n) \label{eq:head}
\end{align}
In \eqref{eq:mhead} and \eqref{eq:head}, the $n$th layer is computed with the projection matrices $\W_{Q,i}^n \in \R^{d_{\model} \times d}$, $\W_{K,i}^n \in \R^{d_{\model} \times d}$, $\W_{V,i}^n \in \R^{d_{\model} \times d}$, and $\W_{O}^n \in \R^{md \times d_{\model}}$.
For all the SANs in the encoder, $\Q$, $\K$, and $\V$ are the same matrices, which are the inputs of SAN. 
The position-wise feed-forward network is a stack of linear layers.

The decoder predicts the probability of the following character from previous output characters and the encoder output $\h$, i.e., $p(y_s|y_1,\dots,y_{s-1},\h)$, similarly to that in LMs.
The character history sequence is converted to character embeddings.
Then, $N_{d}$ decoder layers are applied, followed by linear projection and the Softmax function.
The decoder layer consists of two SANs followed by a position-wise feed-forward network.
The first SAN in each decoder layer applies attention weights to the input character sequence, where the input sequence of the SAN is set as $\Q$, $\K$, and $\V$.
Then the following SAN attends to the entire encoder output sequence by setting $\K$ and $\V$ to be the encoder output $\h$.
By using a mask function as in \cite{vaswani17,van16}, the decoding process is carried out without recurrence; thus, both the encoder and the decoder are efficiently trained in an E2E manner.

\section{Contextual block processing}
\label{sec:block}

\subsection{Block encoding}
\label{ssec:block-enc}
In the applications of real-time ASR, which receive a speech data stream, the recognition should be performed online.
Most of the state-of-the-art systems, such as the bidirectional LSTM (biLSTM) \cite{watanabe17} and Transformer \cite{dong18}, require the entire speech utterance for both encoding and decoding; thus, they are processed only after the end of the utterance, which is not suitable for online processing.
Considering that at least phonetic events occur in the local temporal region, the encoder can be computed block-wise, as in \cite{sperber18, dong19, jaitly2015neural}.
In contrast, the decoder follows a sequential process by its nature, since it emits characters one by one using the output history.
Although it should require at least the part of the encoder output $\h$ corresponding to the processing character, estimating the optimal alignment between the encoder output and the decoder is still difficult, especially without any language dependence.
Therefore, we leave the online processing of the decoder for our future work, and the decoding is performed only after the entire utterance is input.

\begin{figure}[t]
  \centering
  \includegraphics[width=\columnwidth]{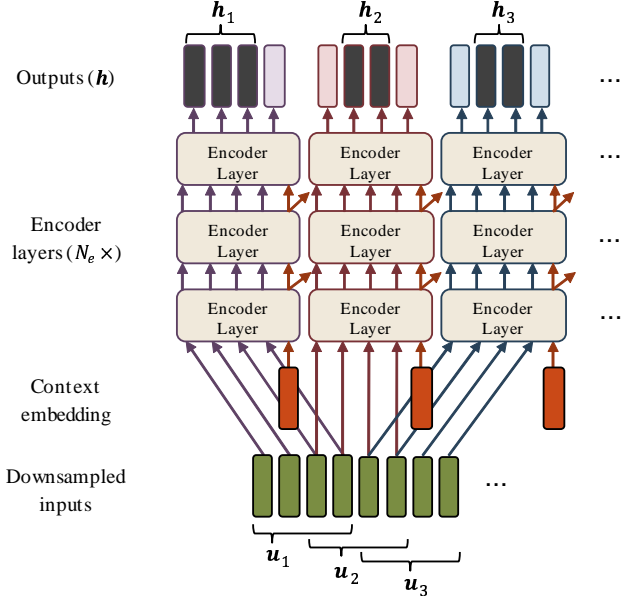}
  \vspace{-0.5cm}
  \caption{The context inheritance mechanism of the encoder.}
  \label{fig:context}
\end{figure}

Denoting downsampled input features, i.e., the output of two convolutional layers, linear projection, and positional encoding, as $\u=(u_1,\dots,u_{T/4})$, the block size as $\Lblock$, and the hop size as $\Lhop$, the $b$th block is processed using input features $u_{t}$ from $t=(b-1) \cdot \Lhop + 1$ to $t=(b-1) \cdot \Lhop + \Lblock$, which is denoted as an $\Lblock$-length subsequence $\u_b$, i.e.,
\begin{equation}
    \label{eq:block_feat}
    \u_b = (u_{(b-1) \cdot \Lhop + 1}, \dots, u_{(b-1) \cdot \Lhop + \Lblock})
\end{equation}
Considering the overlap of the blocks, we utilize only the central $\Lhop$ of the output of each block, which we denote as $\h_b$.
The first frames are included in the first block $\h_1$ and the last frames are in the last block $\h_B$.
When $\Lblock=16$ and $\Lhop=8$, block encoding is performed using 64-frame acoustic features with 32 frames overlap.
The block encoding is depicted in Fig.~\ref{fig:context}.

\subsection{Context inheritance mechanism}
\label{ssec:context}
Global channel, speaker, and linguistic context are also important for local phoneme classification.
To utilize such context information, we propose a new context inheritance mechanism by introducing an additional context embedding vector.
As shown in Fig.~\ref{fig:context}, the context embedding vector is computed in each layer of each block and handed over to the upper layer of the following block.
Thus, the SAN in each layer is applied to the block input sequence using the context embedding vector.

The proposed Transformer with the context embedding vector is straightforwardly extended from the original formulation in Section~\ref{sec:transformer}.
The SAN computations \eqref{eq:mhead} and \eqref{eq:head} of layer $n$ of block $b$ are rewritten as 
\begin{align}
    & \mathrm{MHD}(\Tilde{\Q}_b^n,\Tilde{\K}_b^n,\Tilde{\V}_b^n) = \mathrm{Concat}(\Tilde{\head}_{1},\dots,\Tilde{\head}_{m})\W_O^n,
    \label{eq:bmhead} \\
    & \Tilde{\head}_{i} = \mathrm{Attention}(\Tilde{\Q}_b^n\W_{Q,i}^n,\Tilde{\K}_b^n\W_{K,i}^n,\Tilde{\V}_b^n\W_{V,i}^n), \label{eq:bhead}
\end{align}
where $\Tilde{\mbox{}}$ denotes the augmented variables with the context embedding vector.
Denoting the context embedding vector as $\mathbf{c}_{b}^{n}$, the augmented variables are defined as follows:
\begin{itemize}
    \item In the first layer ($n=1$),  $\Tilde{\Q}_b^1=\Tilde{\K}_b^1=\Tilde{\V}_b^1=[\u_b \ c_b^0]$ is represented as an augmented feature matrix composed of the blocked input $\u_b$ as introduced in \eqref{eq:block_feat} and the additional initial context embedding vector $c_b^0$. 
The initialization of $c_b^0$ is discussed in Section~\ref{sssec:init_emb}.
    \item In the succeeding layers ($n>1$), $\Tilde{\Q}_b^{n} = [\Z_b^{n-1} \ c_{b}^{n-1}]$ and $\Tilde{\K}_b^{n}=\Tilde{\V}_b^n=[\Z_b^{n-1} \ c_{b-1}^{n-1}]$ are similarly augmented with the context embedding vector $c_{b-1}^{n-1}$ in the previous block $(b-1)$ of the previous layer $(n-1)$ and $\Z_b^{n-1}$.
$\Z_b^{n}$ is the output of the $n$th encoder layer of block $b$, which is computed simultaneously with the context embedding vector $c_b^{n}$ as 
\begin{align}
    \Tilde{\Z}_b^{n} &= [\Z_b^{n} \ c_b^{n}] \nonumber \\
    &= \max(0, \Tilde{\Z}_{b,\text{int.}}^{n}\W_1^n + v_1^n)\W_2^n + v_2^n + \Tilde{\Z}_{b,\text{int.}}^{n} \\
    \Tilde{\Z}_{b,\text{int.}}^{n} &= \mathrm{MHD}(\Tilde{\Q}_b^{n},\Tilde{\K}_b^{n},\Tilde{\V}_b^{n}) + \Tilde{\V}_{b}^{n},
\end{align}
where $\W_1^n$, $\W_2^n$, $v_1^n$ and $v_2^n$ are trainable matrices and biases.
\end{itemize}
Note that most of these calculations are closed within block $b$.
Only the context embedding part $c_{b-1}^{n-1}$ carries over the previous block context information.
Therefore, if SAN attends to the context embedding vector $c_{b-1}^{n-1}$, the output of $\mathrm{MHD}(\cdot)$ in \eqref{eq:bmhead} delivers the context information to the succeeding layer as the tilted arrows in Figure~\ref{fig:context}.

This framework enables a deeper layer to hold longer context information.
For example, $c_b^n$ corresponds to an $(n-1)$-block-length context, by expanding the above recursive equation between $c_b^n$ and $c_{b-1}^{n-1}$ (denoted as $f(\cdot)$) as follows:
\begin{align}
    c_b^n &= f(\Z _{b} ^{n-1}, c_{b-1}^{n-1}) = f(\Z _{b} ^{n-1}, f(\Z _{b-1} ^{n-2}, c_{b-2}^{n-2})) \cdots \nonumber\\
    &= g(\Z _{b} ^{n-1}, \dots, \Z _{b-n+2} ^{1}, \u_{b-n+1}).
\end{align}
Thus, our proposed framework can include long context information while keeping the basic block processing.

\subsection{Context embedding initialization}
\label{sssec:init_emb}
For the initial context embedding vector for the first layer, $c_b^{0}$, we propose three types of initialization as follows.
\subsubsection{Positional encoding}
\label{sssec:position}
For the initial context embedding vector, a simple positional encoding process (\ref{eq:position}) is adopted.
The position ($pos$) is rearranged over the blocks, starting from $0$.
\begin{align}
    c_{(b,i)}^{0} = \mathrm{PE}_{(b-1,i)}
\end{align}
For the following layers, only the contextual output of each encoder layer $c_b^{n}$ is handed over.

\subsubsection{Average input}
\label{sssec:average}
We expect the context embedding to inherit global statistics from the precedent blocks.
Therefore, the average of the input sequence for the block $\u_{b}$ is used as the initial context embedding vector because it is a statistic that has already been obtained.
\begin{align}
    c_{b}^{0} = \mathrm{Average}(\u_b)
\end{align}
Positional encoding can also be added to the average to help identify the sequence of blocks.

\subsubsection{Maximum values of input}
\label{sssec:max}
Instead of taking the average, the maximum values are taken along the temporal axis as 
\begin{align}
    c_{b}^{0} = \mathrm{Maxpool}(\u_b).
\end{align}
This can also be combined with the positional encoding.

\subsection{Implementation}
\label{ssec:implementation}
One of the advantages of the Transformer is its efficiency in training.
Since the Transformer is not a recursive network, it can be trained in parallel without waiting for preceding outputs.
Even for the causal process of the decoder, the training is performed at a single time using a mask function as in \cite{vaswani17,van16}.
Our proposed contextual block processing can also be implemented in a similar manner.

\begin{figure}[t]
  \centering
  \includegraphics[width=1.\columnwidth]{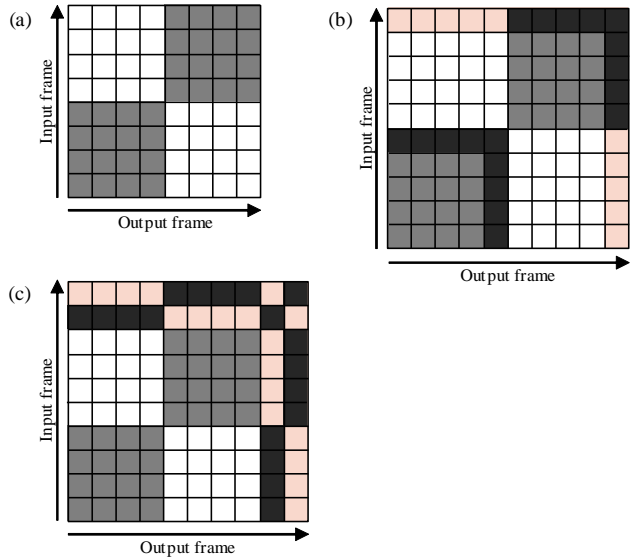}
  \caption{The design of masks used for block processing. (a) is for naive block processing, (b) is an extension for contextual block processing for the first layer, (c) is an alternate form of (b). Colored regions are for context embedding vectors and all the darken regions pass values.}

  \label{fig:mask}
\end{figure}

For the block encoding, the mask shown as (a) in Fig.~\ref{fig:mask} is designed to confine the Softmax and output computation in (\ref{eq:attention}) within a block.
This is an example of $L_{block}=4$, which narrows the frames used for Softmax computation down to 1--4 for the output frame 1--4, and down to 5--8 for the output frame 5--8.
The central frames of each block are extracted after the last layer computation and stacked in the matrix $\h$.
In the case of overlapping, multiple masks are created and applied individually.
For instance, the case of half-overlapping ($\Lhop = \Lblock/2$), two masks with $\Lhop$ frame shifting are used.

Contextual inheritance requires additional vectors for context embedding.
Straightforwardly, the context embedding vectors are inserted after each block, then the mask is extended as (b) in Fig.~\ref{fig:mask} for the first layer, where colored regions correspond to the context embedding vectors.
Thus, for computing both the output of SAN $Z_{b}^1$ and the context embedding vector $c_b^1$, both the input $\u_b$ and the initial context embedding vector $c_b^0$ are used.
Masks for the succeeding layers are designed to attend to the context embedding vectors of the preceding blocks.
However, the insertions of the context embedding vectors complicate the implementation.
Instead, we simply concatenate $B = T/4\Lhop$ frames of context embedding vectors to the end of the input sequence $\u$ as
\begin{align}
    \u_{\mathrm{ext}} = (u_1,\dots,u_{T/4}, c_1^0,\dots,c_B^0).
\end{align}
Then, the mask shown as (c) in Fig.~\ref{fig:mask} is designed for the contextual block processing to include the context embedding vector in the Softmax computation and produce a new context embedding vector.

Technically, in our implementation, each layer normalization is applied to the entire input sequence; thus, in each block, the statistics are shared over all the blocks.
However, we leave this unchanged for efficiency, because our preliminary experiment showed this global normalization was not significantly different from the normalization within each block.
Additionally, in the case of half-overlapping, the context embedding vectors cannot be referred to across the different two masks.
Therefore, each block refers to the context embedding vector of two blocks earlier; thus, $\Tilde{\K}_b^n=\Tilde{\V}_b^n=[\Z_b^{n-1} \ c_{b-2}^{n-1}]$.

\section{Experiments}
\label{sec:experiments}

\subsection{Experimental setup}
\label{ssec:setups}
We carried out experiments using the WSJ and LibriSpeech datasets \cite{panayotov15} for English ASRs.
The models were trained on the SI-284 set and evaluated on the eval92 set for the WSJ, and we used 960 hours of training data and both clean and contaminated test data were evaluated for the LibriSpeech.
Also VoxForge\footnote{VoxForge: \url{http://www.voxforge.org}} Italian corpus and AISHELL-1 Mandarin data \cite{aishell17} are trained and evaluated.
The input acoustic features were 80-dimensional filterbanks and the pitch, extracted with a hop size of 10 ms and a window size of 25 ms, which were normalized with the mean and variance.
For the WSJ setup, the number of output classes was 52, including the 26 letters of the alphabet, space, noise, symbols such as period, an unknown marker, a start-of-sequence (SOS) token, and an end-of-sequence (EOS) token.
Similarly, we used 36 character classes for the VoxForge Italian and 4231 character classes for the AISHELL-1 Mandarin.
For the LibriSpeech, we adopted byte-pair encoding (BPE) subword tokenization \cite{sennrich16}, which ended up with 5000 token classes.

For the training, we utilized multitask learning with CTC loss as in \cite{watanabe17} with a weight of 0.3.
A linear layer was added onto the encoder to project $\h$ to the character probability for the CTC. 
The Transformer models were trained over 100 epochs for WSJ, 50 epochs for LibriSpeech/AISHELL-1, and 300 epochs for VoxForge, with the Adam optimizer and Noam learning rate decay as in \cite{vaswani17}, starting with a learning rate of 5.0, and a mini-batch size of 32.
The parameters of the last 10 epochs were averaged and used for inference.
The encoder had $N_{e}=12$ layers with 2048 units and the decoder had $N_{d}=6$ layers with 2048 units, with both having a dropout rate of 0.1.
For the SAN, the output $\Tilde{\mathbf{Z}}_b^n$ was a 256-dimensional vector ($d_{model}=256$) with four heads ($m=4$).
We trained three types of the Transformer, the baseline Transformer without any block processing, the Transformer with naive block processing as described in Section~\ref{ssec:block-enc}, and the Transformer with our proposed contextual block processing method as in Section~\ref{ssec:context}.
The training was carried out using ESPNet\footnote{ESPNet: \url{https://github.com/espnet/espnet/}} \cite{watanabeespnet} with  the PyTorch\footnote{PyTorch: \url{https://pytorch.org/}} toolkit \cite{paszke17}.

The decoding was performed alongside the CTC, whose probabilities were added with a weight of 0.3 to those of the Transformer.
We performed decoding using a beam search with a beam size of 10.
An external word-level LM was used for rescoring using shallow-fusion \cite{kannan18} with a weight of $1.0$ for WSJ, $0.7$ for LibriSpeech, and $0.3$ for AISHELL-1; this model was a single-layer LSTM with 1000 units.
We did not use an external LM for the VoxForge dataset.

We also trained the biLSTM model \cite{watanabe17} as baselines.
The baseline models consisted of an encoder with a VGG layer, followed by bidirectional LSTM layers and a decoder.
The numbers of the encoder biLSTM layers were six, five, four, and three, with 320, 1024, 320, and 1024 units for WSJ, LibriSpeech, VoxForge, and AISHELL-1 respectively.
The decoders were an LSTM layer with 300 units for WSJ and VoxForge, and two LSTM layers with 1024 units for the rest.
For the online encoding, it is a natural idea to utilize a unidirectional LSTM.
Thus, for a comparison, simple LSTM models, in which bidirectional LSTMs were swapped with unidirectional LSTMs, were trained in the same conditions.

\begin{table}[t]
  \caption{Word error rates (WERs) in the WSJ evaluation task with $\Lblock=16$ and $\Lhop=8$.}
  \label{tab:wsj}
  \centering
  \begin{tabular}{l|c}
    \hline
      & eval92 \\
    \hline\hline
    \multicolumn{2}{l}{Batch encoding}  \\
    \hline
    biLSTM \cite{watanabe17} & 6.7 \\
    Transformer & {\bf 5.0} \\
    \hline
    \multicolumn{2}{l}{Online encoding} \\
    \hline
    LSTM & 8.4 \\
    Block Transformer & 7.5 \\
    Contextual Block Transformer \\
    \ \ ---{\it PE} & 6.0 \\
    \ \ ---{\it Avg. input} & 6.3 \\
    \ \ ---{\it Max input} & 10.9 \\
    \ \ ---{\it PE + Avg. input} & {\bf 5.7} \\
    \ \ ---{\it PE + Max input} & 7.9 \\
    \hline
  \end{tabular}
\end{table}

\subsection{Results}
\label{ssec:results}

\subsubsection{Comparison of recognition performance}
\label{sssec:wer}
We first carried out a word error rate (WER) comparison on the WSJ dataset.
The results are summarized in Table~\ref{tab:wsj}.
For the batch encoding, we observed that the Transformer outperformed the conventional biLSTM model.
It degraded when each biLSTM was swapped with a LSTM.
When we used naive online block processing for the encoding, as in Section~\ref{ssec:block-enc}, the error rate decreased from that of the LSTM model.
For our proposed contextual block processing method, various context embedding vector initializations were tested, initialization with positional encoding as in Section~\ref{sssec:position} ({\it PE}), with the average input as in Section~\ref{sssec:average} ({\it Avg. input}), with the maximum values of inputs as in Section~\ref{sssec:max} ({\it Max input}), and their combinations ({\it PE + Avg. input} and {\it PE + Max input}).
The proposed contextual block processing methods using {\it PE} and {\it Avg. input} improved the accuracy significantly, among which the combination ({\it PE + Avg. input}) achieved the best result.

We also applied each model to the LibriSpeech, VoxForge Itallian, and AISHELL-1 Mandarin datasets.
Only the {\it PE + Avg. input} initialization was adopted for the context embedding vector.
The results shown in Table~\ref{tab:librispeech} have a similar tendency to those for the WSJ dataset.
This indicates that our proposed context inheritance mechanism is consistently useful to leverage the global context information.

\begin{table}[t]
  \caption{WERs/CERs for the LibriSpeech, VoxForge Italian, and AISHELL-1 Mandarin datasets ($\Lblock=16$, $\Lhop=8$).}
  \label{tab:librispeech}
  \centering
  \begin{tabular}{l|c|c|c|c}
    \hline
    &\multicolumn{2}{c|}{LibriSpeech} & VoxForge &  AISHELL \\
    &\multicolumn{2}{c|}{(WER)} & (CER) &  (CER) \\
    & clean & other & &\\
    \hline\hline
    \multicolumn{5}{l}{Batch encoding}  \\
    \hline
    biLSTM \cite{watanabe17}  & {\bf 4.2} & 13.1 & 10.5 & 9.2\\
    Transformer & 4.5 & {\bf 11.2} & {\bf 9.3} & {\bf 6.4}\\
    \hline
    \multicolumn{5}{l}{Online encoding} \\
    \hline
    LSTM & 5.3 & 16.1 & 14.6 & 11.8 \\
    Block & 4.8 & 13.2 & 11.5 & 7.8 \\
    Contextual Block &&&&\\
    \ \ ---{\it PE + Avg. input} & {\bf 4.6} & {\bf 13.1} & {\bf 10.3} & {\bf 7.6}\\
    \hline
  \end{tabular}
\end{table}

\subsubsection{Comparison of block size}
\label{sssec:blocksize}
\begin{figure}[t]
  \centering
  \includegraphics[width=0.9\columnwidth]{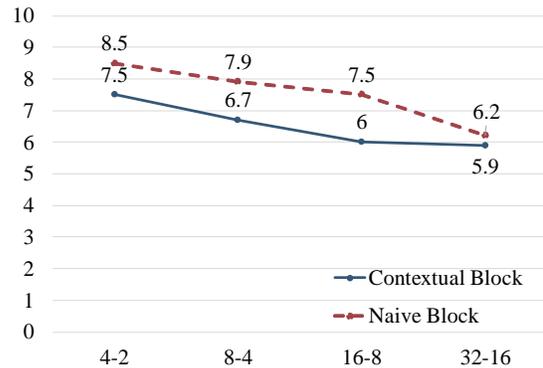}
  \caption{The WERs in the WSJ evaluation task for various block sizes ($\Lblock$ -- $\Lhop$).}
  \label{fig:wer}
\end{figure}

We also evaluated WERs for various block sizes ($\Lblock$), i.e., 4, 8, 16, and 32, for the naive block Transformer and contextual block Transformer on the WSJ dataset.
The block processing was carried out in the half-overlapping manner; thus, $\Lhop = \Lblock/2$.
The results are shown in Fig.~\ref{fig:wer}.
As the block size increased, the performance improved in both types of block processing.
The proposed processing method consistently had better performance, especially with small block sizes.
This result also indicates that a block size of 16 is sufficient for contextual block processing.
Interestingly, this result also shows that a block size of 32 is sufficient to acquire certain context information, where the contextual block processing method improved only a small amount.

\subsubsection{Interpretation of attention weight}
\label{sssec:attention}
\begin{figure}[t]
  \centering
  \includegraphics[width=1\columnwidth]{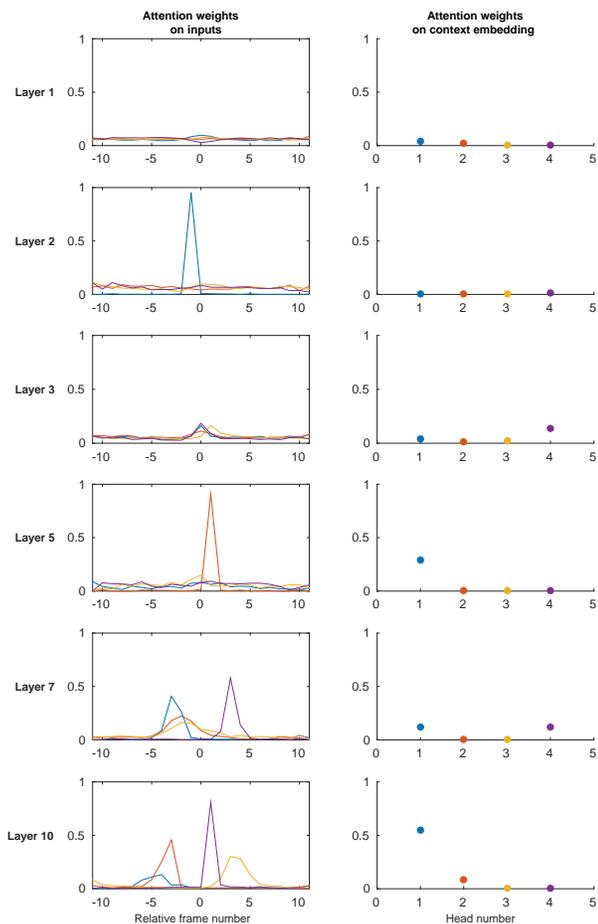}
  \vspace{-1.5cm}
  \caption{Attention weights used for computing outputs of self-attention network (SAN) over a WSJ utterance sample in each layer, applied to the inputs (left column) and to the context embedding vector (right column).  Each color corresponds to the same head. }
  \label{fig:attention}
\end{figure}

We also looked at how the attention works in the proposed context inheritance mechanism.
We sampled an utterance in the WSJ evaluation data to compute the statistics of the attention weights using the Softmax in (\ref{eq:attention}).
Fig.~\ref{fig:attention} shows the attention weights in the layers that we considered, which are used for computing the outputs of encoder layer $\Z_{b}^{n}$, applied to the inputs $\mathbf{V}_b^n = \Z_b^{n-1}$ (left column) and to the context embedding vector $c_{b-2}^{n-1}$ (right column).  
Each color corresponds to the same head.

Looking at the left column, in the first layer, the attention weights tended to evenly attend to the input sequence, while the context embedding vector was not attended to.
In deeper layers, the attention weights started to develop peaks in the center, and the weights for the context embedding vector (right column) started to increase from the third layer.
This indicates that the deeper layers rely on the context information more.
For instance, the first head of layer 5 (blue color) did not strongly attend to the inputs of the SAN, whereas it attended to the context embedding vector with a weight of 0.3.
Interestingly, in the seventh and tenth layer, the first and second heads (blue and orange) used the inputs of a few frames earlier to encode each frame, with and without the context information, and the fourth head (purple) attended to the future.
The first head of layer 10 integrated information over the contexts of nine blocks, which consists of 576 frames (5.76 s), with more attention weight than that of the input.

\section{Conclusion}
\label{sec:conclusion}
We have proposed a new block processing method of the Transformer encoder by introducing a context-aware inheritance mechanism.
An additional context embedding vector handed over from the previously processed block helped to encode not only local acoustic information but also global linguistic/channel/speaker attributes.
We extended a mask technique to realize efficient training with the context inheritance.
Evaluations of the WSJ, LibriSpeech, VoxForge Italian, and AISHELL-1 Mandarin speech recognition datasets showed that our proposed contextual block processing method outperformed naive block processing consistently.
We also analyzed the attention weight tendency of each layer to interpret the behavior of the added contextual inheritance mechanism.

In this study, we used the original Transformer for decoding, which will be alternated with online processing in our future work.
We will also investigate the computation cost and delay with the online decoding implementation in a streaming recognition scenario.


\bibliographystyle{IEEEbib}
\bibliography{mybib}

\end{document}